%% file: paper.tex
\documentclass[10pt,conference]{IEEEtran}
\IEEEoverridecommandlockouts

\usepackage{balance}
\usepackage[percent]{overpic}
\usepackage{tikz}
\usepackage{hyperref}
\usepackage{cite}
\usepackage{amsmath,amssymb,amsfonts}
\usepackage{algorithmic}
\usepackage{graphicx}
\usepackage{textcomp}
\usepackage{xcolor}
\def\BibTeX{{\rm B\kern-.05em{\sc i\kern-.025em b}\kern-.08em
    T\kern-.1667em\lower.7ex\hbox{E}\kern-.125emX}}

\usepackage{xspace}
\newcommand{\tool}{\textsc{TSBugs Explorer}\xspace}

\definecolor{mycommentcolor}{RGB}{100,100,255}  

\begin{document}

\title{TSBugs Explorer: Exploring and Benchmarking TypeScript Bugs}

\author{\IEEEauthorblockN{TianYi Tang}
\IEEEauthorblockA{\textit{School of Computing Science} \\
\textit{Simon Fraser University}\\
Burnaby, BC, Canada \\
tta89@sfu.ca}
\and
\IEEEauthorblockN{Saba Alimadadi}
\IEEEauthorblockA{\textit{School of Computing Science} \\
\textit{Simon Fraser University}\\
Burnaby, BC, Canada \\
saba@sfu.ca}
}

\maketitle

\input{src/abstract}


\input{src/introduction}
\input{src/explorer}

\input{src/scenarios}

\input{src/related}

\input{src/availability}
\input{src/conclusion}
\input{src/ack}

\balance

\bibliographystyle{IEEEtran}
\bibliography{IEEEabrv,refs}

\end{document}

%% file: src/abstract.tex
\begin{abstract}
Curated datasets of real-world software bugs support empirical
research in software testing, debugging, and automated program
repair. However, exploring heterogeneous bug collections,
comparing fault characteristics across projects, and constructing
research-specific subsets often require manual inspection and
custom scripts. We present \tool, an interactive visual analytics
platform built upon our previously-published dataset of 633 bugs
from 16 open-source TypeScript repositories, classified into
11 fault categories. The platform integrates bug records,
repository characteristics, dependency information, and taxonomy
guidance through coordinated visualizations, interactive filtering,
and detailed bug inspection. It supports alternative aggregation
and normalization of fault distributions, pairwise repository
comparisons, and analysis of differences in observed fault profiles.
An integrated benchmark builder allows researchers to select
bug subsets using multiple criteria, inspect their composition,
and export records in CSV or JSON format.
We demonstrate the platform through three research workflows:
fault characterization, cross-repository comparison, and
task-specific benchmark construction. Functional validation
by two authors confirmed selection correctness, visualization
consistency, export integrity, and reproducibility across
the tested workflows.
By making the existing corpus easier to explore,
compare, and reuse, \tool facilitates systematic,
research-specific selection and export of TypeScript
bug records.
\end{abstract}

%% file: src/introduction.tex
\section{Introduction}
\label{sec:introduction}

Curated datasets of real-world software bugs support empirical
research and the evaluation of testing, debugging, fault
localization, and automated program repair techniques
~\cite{just2014defects4j,gyimesi2020bugsjs}.
However, different research questions often require different
subsets of bugs, selected by fault category, application domain,
or repository. Effective dataset reuse therefore requires
understanding the corpus composition, identifying relevant
instances, and assessing their suitability for a study.
Such selection decisions can influence empirical findings
~\cite{bird2009bias}, yet exploring heterogeneous bug
collections and constructing research-specific subsets often
involves manual inspection and custom data-processing scripts.

In previous work~\cite{tang2026logic}, we introduced a
curated dataset of 633 bugs from 16 open-source TypeScript
repositories, classified into 11 fault categories. We
characterized fault distributions and examined their
relationships with repository characteristics, application
domains, and dependency structures. Although the dataset and
analysis artifacts were publicly released, researchers lacked
an integrated interface for inspecting individual bugs,
comparing repositories, and selecting subsets for further
study. We address this limitation with \tool, a web-based
visual analytics platform that brings together bug records,
fault classifications, repository characteristics, dependency
metadata, and taxonomy guidance.

\tool combines coordinated visualizations and interactive
filtering with searchable bug inspection, normalized
cross-repository comparisons, and taxonomy guidance.
Its benchmark builder enables researchers to select bug
records using repository, fault-category, application-domain,
and changed-file criteria, inspect the resulting distributions,
and export the selected metadata in CSV or JSON format.
By connecting exploratory analysis with configurable subset
construction, the platform facilitates the systematic
selection and reuse of TypeScript bug records. This paper
describes the platform, illustrates three research workflows,
and reports its functional validation and limitations.

%% file: src/explorer.tex
\section{TSBugs Explorer: Dataset and Platform}
\label{sec:explorer}

\subsection{Dataset and Implementation}

\tool builds on our previously-published TypeScript bug
dataset~\cite{tang2026logic}, comprising 633 bug instances
from 16 open-source repositories, classified into 11 fault
categories. It integrates bug records, repository characteristics,
application domains, dependency metadata, and taxonomy
definitions. Bug records contain repository identifiers, fault
categories, changed-file counts and paths, and links to fixing
commits. Repository metadata includes code size, activity,
popularity, and development technologies.

\tool is implemented as a client-side web application using
JavaScript and D3.js\footnote{\url{https://d3js.org/}}
for interactive visualizations. Data is
stored in separate JSON files for bugs, repositories, domains,
dependencies, and taxonomy definitions, separating dataset
content from application logic. Filtering, aggregation,
visualization, and export are performed in the browser,
without a dedicated backend or database. The platform
provides five views: Dashboard, Bug Explorer, Compare Projects,
Benchmark Builder, and Taxonomy.

\subsection{Interactive Exploration and Inspection}

The \emph{Dashboard} provides interactive visualizations
of fault-category distributions, repository coverage,
application domains, and repository characteristics
(\autoref{fig:dashboard}). Its filtering and search toolbar
(\autoref{fig:dashboard}(a)) supports selection by repository,
fault category, and application domain, as well as keyword
search across bug records, repositories, and affected files.
Filters are coordinated across Dashboard visualizations
and the Bug Explorer, and can also be applied through direct
interaction with chart elements. Active criteria appear as
removable filter chips, allowing users to combine or
reverse selections.

The category distribution (\autoref{fig:dashboard}(b))
offers two aggregation modes: \emph{Sample Count}, reporting
the number of collected bugs per category, and
\emph{Project-Weighted}, averaging category proportions
across repositories so that each contributes equally,
regardless of its sampled bug count. These modes distinguish
the composition of the collected sample from the average
fault-category profile across repositories.

The \emph{Fault Landscape Heatmap}
(\autoref{fig:dashboard}(d)) displays fault-category
distributions across repositories, with rows and columns
adapting to active filters. Its within-project percentage
mode complements raw counts for comparing category
composition across repositories. Other visualizations
show repository coverage (\autoref{fig:dashboard}(c)),
application-domain distributions (\autoref{fig:dashboard}(e)),
and repository size and popularity, with markers indicating
sampled bug counts (\autoref{fig:dashboard}(f)).

The \emph{Bug Explorer} supports keyword search,
attribute-based filtering, and inspection of individual
bug records (\autoref{fig:workflows}(a)). Each record
provides its classification, affected file paths, fixing
commit, repository characteristics, and related bugs.
Shared filters connect aggregate patterns to individual
bug instances.

The \emph{Compare Projects} view enables pairwise repository
comparison through normalized fault-category distributions,
repository metrics, and dependency/tooling profiles
(\autoref{fig:workflows}(b)). Its \emph{Category Difference}
visualization highlights percentage-point differences
between repository fault profiles. The \emph{Taxonomy} view
provides category definitions, classification guidance,
commonly confused categories, and representative examples.
Together, these views support exploration at the dataset,
repository, and individual-bug levels.

\input{figures/figure1}

\subsection{Benchmark Construction and Export}

The \emph{Benchmark Builder} enables researchers to construct
research-specific bug subsets by combining repository,
fault-category, application-domain, and minimum changed-file
criteria (\autoref{fig:workflows}(c)). Searchable multi-selection
controls and Select All and Clear actions support iterative
refinement of selection criteria. The interface dynamically
updates the matching bug count and displays category and
repository distributions, allowing researchers to examine
subset composition before export.

Selected records can be exported in CSV or JSON format,
preserving bug identifiers, classifications, repository
information, changed-file metadata, and fixing-commit links.
JSON export retains the corresponding bug objects, while
a textual summary describes the selected subset. These
facilities support the preparation of task-specific bug
collections for subsequent empirical studies. The current
implementation exports bug metadata rather than reconstructing
executable buggy and fixed program versions.

%% file: figures/figure1.tex
\begin{figure}[t]
    \centering
    \begin{overpic}[
        clip,
        trim={2.7cm 2.9cm 4.4cm 2.6cm},
        width=\columnwidth
    ]{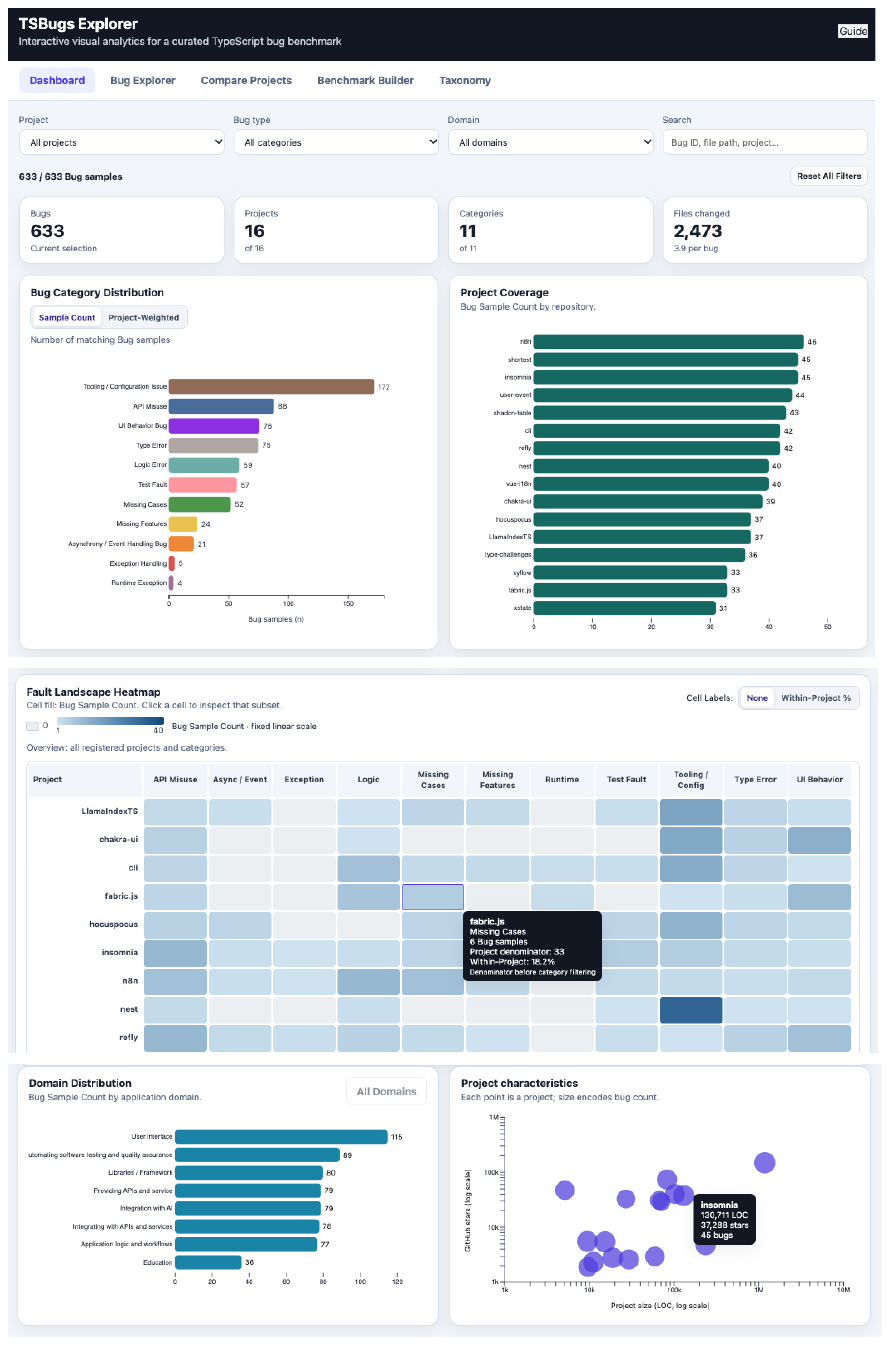}

        \put(60,89){\tikz\node[
            circle, fill=white, draw=black,
            inner sep=1.5pt, font=\bfseries\scriptsize
        ] {(a)};}

        \put(1,52){\tikz\node[
            circle, fill=white, draw=black,
            inner sep=1.5pt, font=\bfseries\scriptsize
        ] {(b)};}

        \put(33,52){\tikz\node[
            circle, fill=white, draw=black,
            inner sep=1.5pt, font=\bfseries\scriptsize
        ] {(c)};}

        \put(1,21){\tikz\node[
            circle, fill=white, draw=black,
            inner sep=1.5pt, font=\bfseries\scriptsize
        ] {(d)};}

        \put(1,1){\tikz\node[
            circle, fill=white, draw=black,
            inner sep=1.5pt, font=\bfseries\scriptsize
        ] {(e)};}

        \put(33,1){\tikz\node[
            circle, fill=white, draw=black,
            inner sep=1.5pt, font=\bfseries\scriptsize
        ] {(f)};}

    \end{overpic}
    \caption{Interactive visualizations in \tool:
    (a) filtering and search toolbar for
repository, fault-category, domain, and keyword queries;
    (b) fault-category distribution with sample-count and project-weighted modes;
    (c) repository coverage;
    (d) fault landscape heatmap showing category distributions
    across repositories;
    (e) application-domain distribution; and
    (f) repository characteristics.}
    \label{fig:dashboard}
\end{figure}

%% file: src/scenarios.tex
\section{Research Workflows and Preliminary Evaluation}
\label{sec:workflows}

We illustrate three research workflows supported by \tool:
characterizing fault distributions, comparing repository
fault profiles, and constructing task-specific bug subsets.

\subsection{Research Workflows}

\textbf{Scenario 1: Characterizing fault distributions.}
Consider a researcher investigating tooling and configuration
faults in TypeScript projects. Our previous study identified
172 such instances (27.2\% of 633 bugs)~\cite{tang2026logic}.
Using the Dashboard (\autoref{fig:dashboard}(b)), the
researcher compares category counts with project-weighted
proportions to distinguish the composition of the collected
sample from the average repository-level profile.

The researcher then examines the Fault Landscape Heatmap
(\autoref{fig:dashboard}(d)), switching between counts and
within-project percentages to compare category distributions
across repositories. Coordinated filters allow refinement
by category, repository, and domain, while the Bug Explorer
(\autoref{fig:workflows}(a)) provides access to matching
records, affected files, and fixing commits. This workflow
connects aggregate patterns with individual bug instances.

\textbf{Scenario 2: Comparing repository fault profiles.}
A researcher comparing two TypeScript repositories selects
them in the Compare Projects view (\autoref{fig:workflows}(b)).
Normalized category distributions account for differences
in sampled bug counts, while the Category Difference
visualization highlights percentage-point contrasts.
Repository characteristics and dependency/tooling profiles
provide context for interpreting the observed differences.

The researcher can investigate contrasting categories
through the Taxonomy and Bug Explorer views, supporting
hypothesis generation and targeted case selection.
These comparisons describe the sampled bugs and do not
establish causal relationships or ecosystem-wide prevalence.

\textbf{Scenario 3: Constructing a task-specific benchmark.}
Consider a researcher selecting candidate bugs for evaluating
TypeScript testing or debugging techniques. Using the
Benchmark Builder (\autoref{fig:workflows}(c)), the researcher
selects four fault categories (Asynchrony/Event Handling,
API Misuse, Logic Error, and Type Error) and can further
restrict the selection by repository, application domain,
or minimum changed-file count.

The interface previews matching bug counts and category
and repository distributions, allowing the researcher to
assess subset composition before exporting records in CSV
or JSON format. The exports preserve bug identifiers,
classifications, changed-file metadata, and fixing-commit
links, supporting subsequent analysis and retrieval of
source artifacts. The resulting subset contains bug metadata;
reconstructing executable buggy versions, environments,
and tests remains a separate task.

\input{figures/figure2}

\subsection{Preliminary Evaluation}
\label{sec:evaluation}

We conducted a structured functional validation of \tool
covering data integrity, selection correctness, visualization
consistency, export integrity, and workflow reproducibility.
Two authors performed predefined checks, independently
cross-validating selected results and operations.

\textbf{Data integrity and selection correctness.}
We compared category distributions with the published
dataset and sampled bug records from every repository and
fault category against source JSON data. We tested
repository, category, domain, and minimum changed-file
filters individually and in combination, comparing selected
identifiers and counts with independently derived reference
sets. We also checked selection/reset behavior and bug details.

\textbf{Visualization consistency.}
We checked category counts, heatmap values, and repository/domain
charts against source data, independently recalculating
project-weighted proportions and signed cross-repository
percentage-point differences. We also verified coordinated
filtering, navigation, filter chips, and reset behaviour.

\textbf{Export integrity.}
We checked CSV and JSON exports against selected records,
including identifiers, counts, classifications, changed-file
metadata, and fixing-commit references. Repeated exports
and zero-result selections were also tested.

\textbf{Workflow reproducibility and availability.}
We reproduced the three research scenarios in
\autoref{sec:workflows}, including the benchmark subset
in \autoref{fig:workflows}(c). We verified all five views,
hosted and local deployments, and documentation and
artifact availability.

All checks passed for the tested cases following
correction of identified data inconsistencies.
These results provide evidence of functional correctness
and workflow reproducibility, rather than exhaustive
verification.

%% file: figures/figure2.tex
\begin{figure}[!htbp]
    \centering

    \begin{overpic}[
        width=\columnwidth,
        clip,
        trim={0cm 14cm 0cm 0cm}
    ]{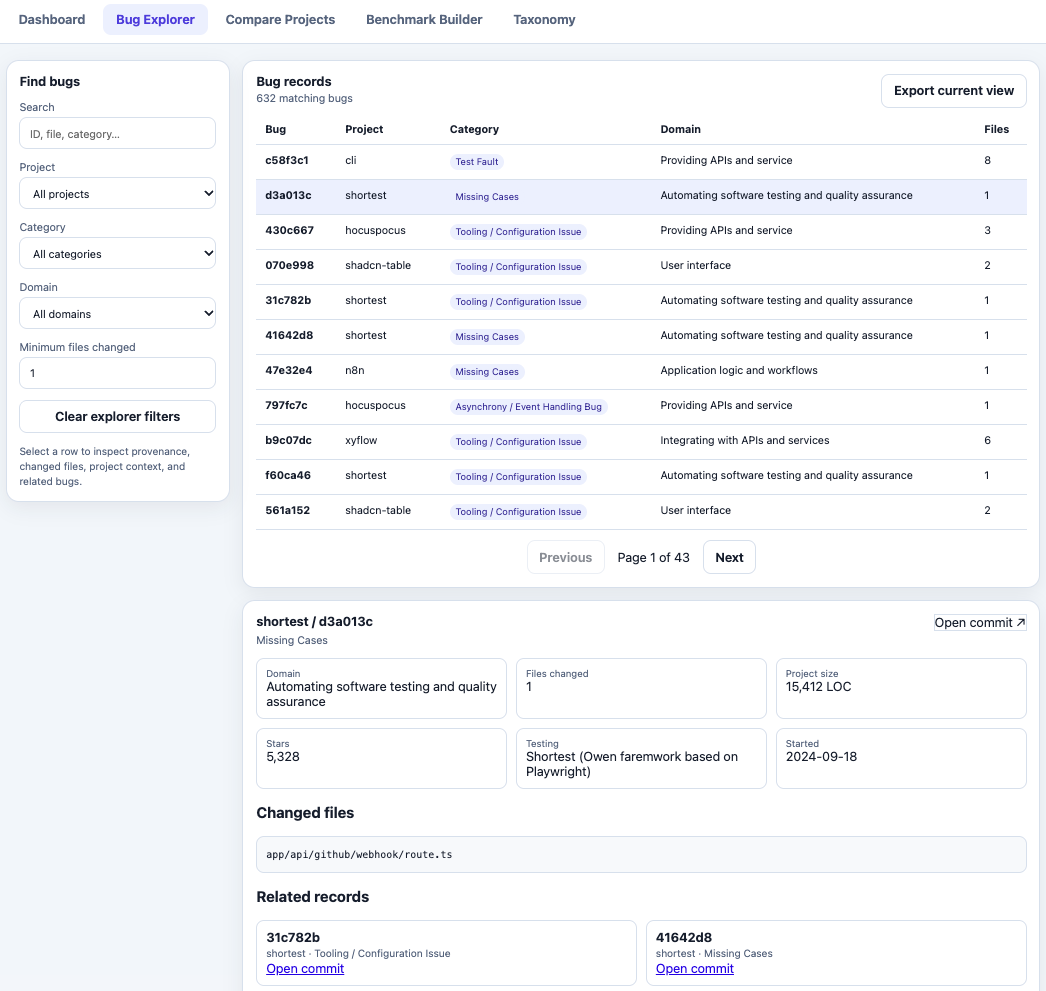}
        \put(92,50){\tikz\node[
            circle, fill=white, draw=black,
            inner sep=1.5pt, font=\bfseries\scriptsize
        ] {(a)};}
    \end{overpic}

    \vspace{3pt}

    \begin{overpic}[
        width=\columnwidth,
        clip,
        trim={0cm 0cm 0cm 0cm}
    ]{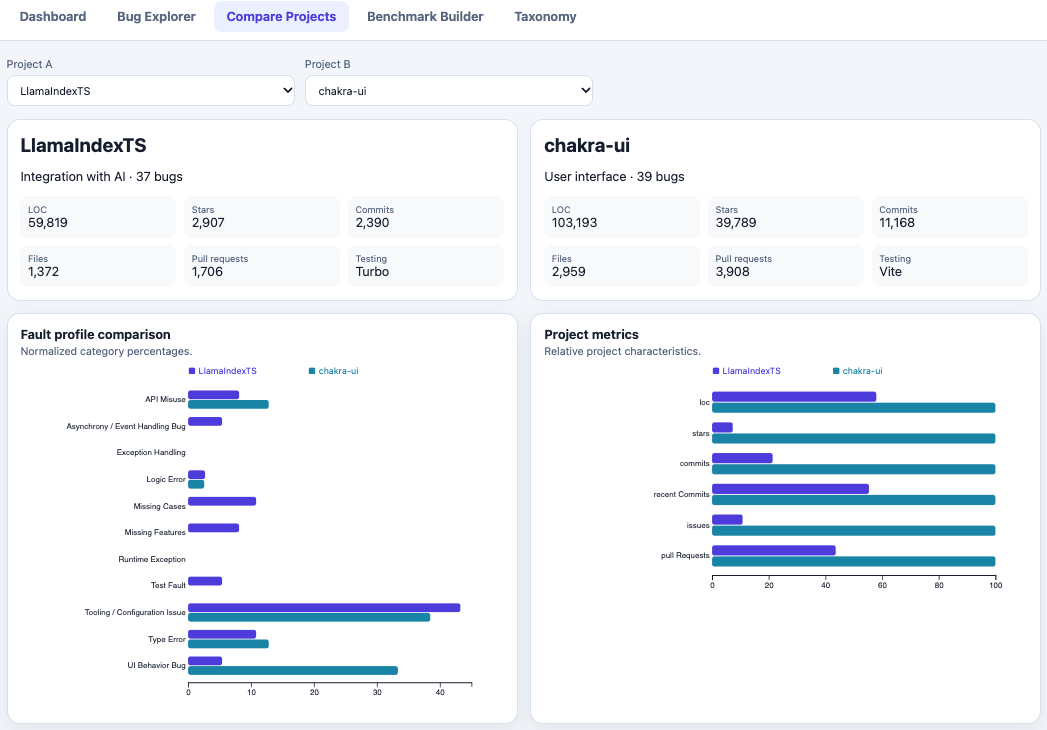}
        \put(92,63){\tikz\node[
            circle, fill=white, draw=black,
            inner sep=1.5pt, font=\bfseries\scriptsize
        ] {(b)};}
    \end{overpic}

    \vspace{-2pt}

    \begin{overpic}[
        width=\columnwidth,
        clip,
        trim={0cm 0cm 0cm 0cm}
    ]{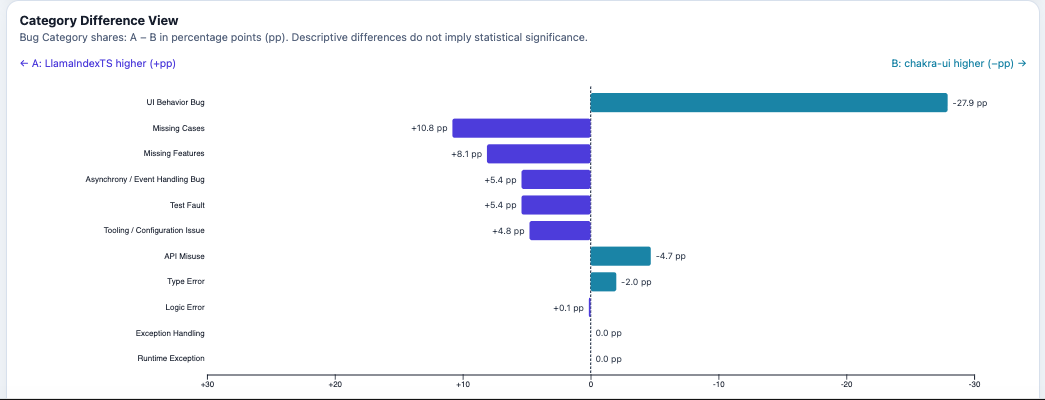}

    \end{overpic}

    \vspace{3pt}
    
    \begin{overpic}[
        width=\columnwidth,
        clip,
        trim={0cm 1.5cm 0cm 0cm}
    ]{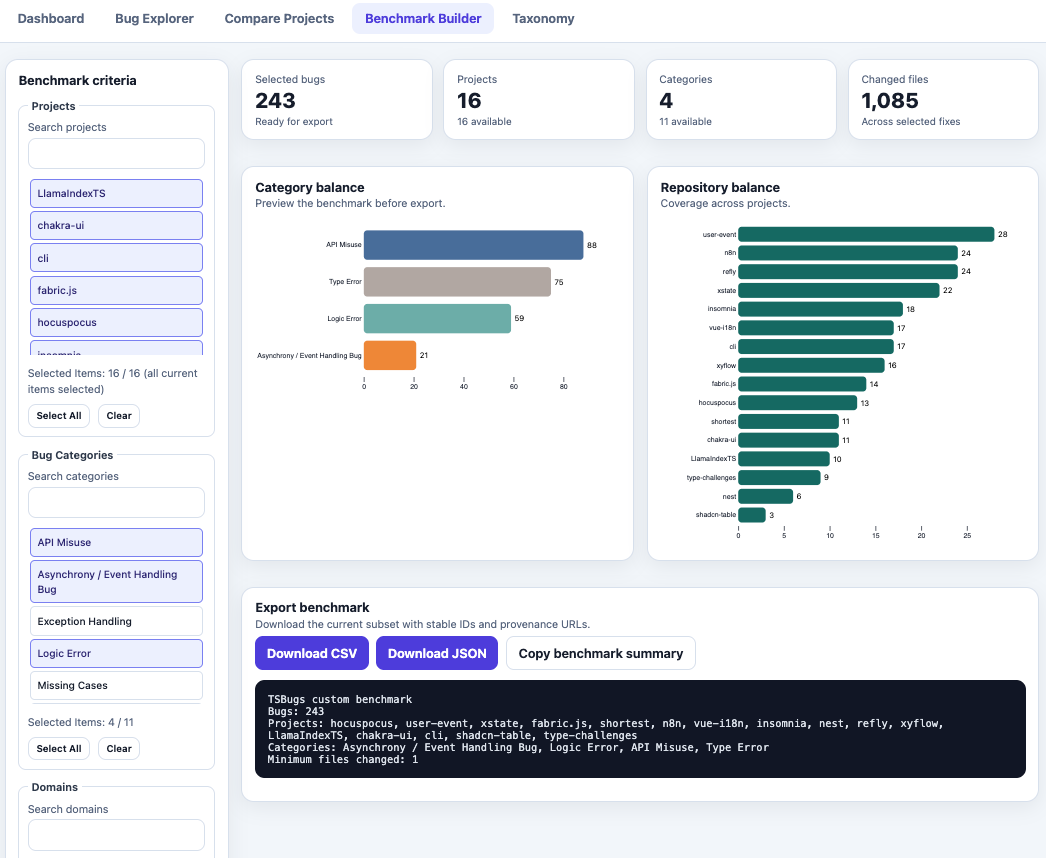}
        \put(92,71){\tikz\node[
            circle, fill=white, draw=black,
            inner sep=1.5pt, font=\bfseries\scriptsize
        ] {(c)};}
    \end{overpic}

    \caption{Selected interfaces supporting research workflows in \tool
    (a) Bug Explorer for searching and inspecting individual
    bug records;
    (b) Compare Projects for examining normalized fault-category
    distributions and differences between repositories; and
    (c) Benchmark Builder for selecting customized bug subsets,
    previewing their composition, and exporting records
    in CSV or JSON format.}
    \label{fig:workflows}
\end{figure}

%% file: src/related.tex
\section{Related Work}
\label{sec:related-work}

\noindent\textbf{Bug Datasets and Benchmarks.}
Curated benchmarks such as Defects4J~\cite{just2014defects4j},
BugsJS~\cite{gyimesi2020bugsjs}, and
BugsInPy~\cite{widyasari2020bugsinpy} provide real-world
bugs for evaluating software testing, debugging, and repair
techniques in Java, JavaScript, and Python, respectively.
Studies of fault taxonomies~\cite{Catolino_2019} and
dataset selection bias~\cite{bird2009bias} further
highlight the importance of understanding benchmark
composition. Our previous work~\cite{tang2026logic}
introduced a curated TypeScript bug corpus and examined
its fault categories in relation to repository characteristics,
application domains, and dependencies. \tool builds on
this corpus to support interactive exploration and
research-specific subset selection rather than introducing
a new bug dataset.

\noindent\textbf{JavaScript and TypeScript Bug Research.}
Prior studies investigated client-side JavaScript
faults~\cite{Ocariza_2013}, concurrency bugs in
Node.js~\cite{Wang_2017}, and performance
bugs~\cite{selakovic2016performance}. Other work
examined the relationship between TypeScript's type
system and software defects~\cite{Gao_2017,Bogner_2022}.
These studies provide empirical evidence about particular
fault mechanisms or language characteristics, whereas
\tool facilitates exploration of a previously classified
TypeScript corpus across fault categories, repositories,
and application domains.

\noindent\textbf{Bug Exploration and Repository Analytics.}
Beyond its benchmark and taxonomy, BugsJS provides a
web-based interface for inspecting bug information and
an API for accessing buggy and fixed versions and
executing tests~\cite{gyimesi2020bugsjs}. General-purpose
repository analysis platforms, including
Boa~\cite{dyer2015boa} and
GrimoireLab~\cite{duenas2021grimoirelab}, support
mining and visualization of software development data.
In contrast, \tool focuses on interactive analysis of
fault-category distributions within a curated TypeScript
corpus, combining coordinated filtering, normalized
cross-repository comparisons, record-level inspection,
and configurable metadata export. Unlike executable
bug benchmarks, its subset builder exports selected
bug records rather than reconstructed program
environments.

%% file: src/availability.tex
\section{Availability, Limitations, and Future Work}
\label{sec:availability}

\noindent\textbf{Availability and Reuse.}
\tool is publicly accessible through a
\href{https://tsbugs-explorer.vercel.app/}{\underline{hosted instance}},\footnote{\url{https://tsbugs-explorer.vercel.app/}}
with its source code and application data available in a
\href{https://github.com/SEatSFU/TSBugsExplorer}{\underline{public GitHub repository}}.\footnote{\url{https://github.com/SEatSFU/TSBugsExplorer}}
A versioned release is permanently archived on
Zenodo~\cite{tsbugs_explorer_2026}.
The underlying TypeScript bug dataset and original analysis
artifacts are maintained in a
\href{https://github.com/SEatSFU/tsbugs}{\underline{separate repository}}
associated with our previous study~\cite{tang2026logic}.
The application can be deployed locally using a static web
server, without a dedicated backend or database.
Its modular JSON files separate bug records, repository
metadata, application domains, dependencies, and taxonomy
information from the visualization logic. The repository
provides documentation for local setup, data organization,
and maintenance. Researchers can explore the corpus through
the hosted or local interface and export selected bug
records in CSV or JSON format.

\noindent\textbf{Limitations and Future Work.}
The platform inherits the sampling and classification
limitations of its underlying dataset, which covers
16 selected TypeScript repositories. Its visualizations
and comparisons characterize the collected bugs rather
than establishing ecosystem-wide fault prevalence or
causal relationships. The Benchmark Builder exports
bug metadata but does not reconstruct executable buggy
and fixed versions, dependencies, or test environments.
Selection criteria are not currently exported as
machine-readable specifications, so users must retain
their criteria separately to reproduce a subset.
Although the separation of data and visualization logic
provides a basis for adapting the platform to other
bug corpora, such adaptation has not been evaluated.
Future work includes expanding the corpus, supporting
reusable selection specifications and configurable
data schemas and taxonomies, and integrating executable
bug environments where feasible.

%% file: src/conclusion.tex
\section{Conclusion}
\label{sec:conclusion}

We presented \tool, an interactive platform for exploring
and reusing a curated TypeScript bug dataset. By integrating
fault distributions, cross-repository comparisons, bug
inspection, and configurable subset export, it extends
the utility of the published corpus beyond static analyses
and manual data processing. The platform supports
research-specific benchmark construction and provides
a reusable resource for empirical studies of TypeScript
reliability, testing, and debugging.

%% file: src/ack.tex
\section*{Acknowledgment}
OpenAI's ChatGPT was used to assist with manuscript
drafting and editing and to generate portions of the initial
software implementation. The authors reviewed, revised,
and refined the AI-assisted outputs and remain responsible
for the final manuscript, software, and reported results.